\documentclass[12pt]{article}
\usepackage{graphicx}
\usepackage{hyperref}

\def \br{{\cal B}}

\def \bd{B^0}

\def \bea{\begin{eqnarray}}
\def \beq{\begin{equation}}

\def \eea{\end{eqnarray}}
\def \eeq{\end{equation}}

\def \lsim{\roughly<}

\def \roughly#1{\mathrel{\raise.3ex\hbox
{$#1$\kern-.75em\lower1ex\hbox{$\sim$}}}}
\def \s{\sqrt{2}}

\begin{document}

\rightline{TECHNION-PH-12-18}
\rightline{UdeM-GPP-TH-12-215}
\rightline{EFI 12-32}
\rightline{November 2012}

\bigskip
\centerline{
\bf\boldmath
RESCATTERING CONTRIBUTIONS TO RARE $B$-MESON DECAYS}
\bigskip

\centerline{Michael Gronau}
\centerline{\it Physics Department, Technion -- Israel Institute of Technology}
\centerline{\it Haifa 3200, Israel}
\medskip

\centerline{David London}
\centerline{\it Physique des Particules, Universit\'e de Montr\'eal}
\centerline{\it C.P. 6128, succ.\ centre-ville, Montr\'eal, QC, Canada H3C 3J7}
\medskip
\centerline{Jonathan L. Rosner}
\centerline{\it Enrico Fermi Institute and Department of Physics}
\centerline{\it University of Chicago, 5620 S. Ellis Avenue, Chicago, IL 60637}
\bigskip

\begin{quote}
Several $B$ 
and $B_s$ decays have been observed which have been cited as evidence for
exchange ($E$), penguin annihilation ($PA$) and annihilation ($A$) processes,
 such as $\bar b d \to \bar u u$, $\bar b s \to \bar u u$ and $\bar b u \to 
 W^* \to \bar c s$, respectively.  These amplitudes are normally thought to be 
 suppressed, as they involve the spectator quark in the weak interaction and thus 
 should be proportional to the $B$-meson decay constant $f_B$. However, as 
 pointed out a number of years ago, they can also be generated by rescattering 
 from processes whose amplitudes do not involve $f_B$, such as color-favored 
 tree amplitudes. In this paper we investigate a number of
processes such as $B^0 \to K^+ K^-$, $B_s \to \pi^+ \pi^-$, and $B^+ \to D_s^+
\phi$, and identify promising states from which they can be generated by
rescattering.  We find that $E$ and $PA$-type processes are characterized
respectively by amplitudes ranging from 5\% to 10\% and from 15\% to 
20\% with respect to the largest amplitude from which they can rescatter.  
Based on this regularity, using approximate flavor SU(3)
symmetry in some cases and time-reversal invariance in others, we predict 
the branching fractions for a large number of as-yet-unseen $B$ and $B_s$ 
decays in an extensive range from order $10^{-9}$ to $10^{-4}$.
\end{quote}

\leftline{PACS numbers: 11.30.Hv, 12.15.Ji, 13.25.Hw, 14.40.Nd}
\bigskip

\section{Introduction}

\begin{figure}
\includegraphics[width=0.9\textwidth]{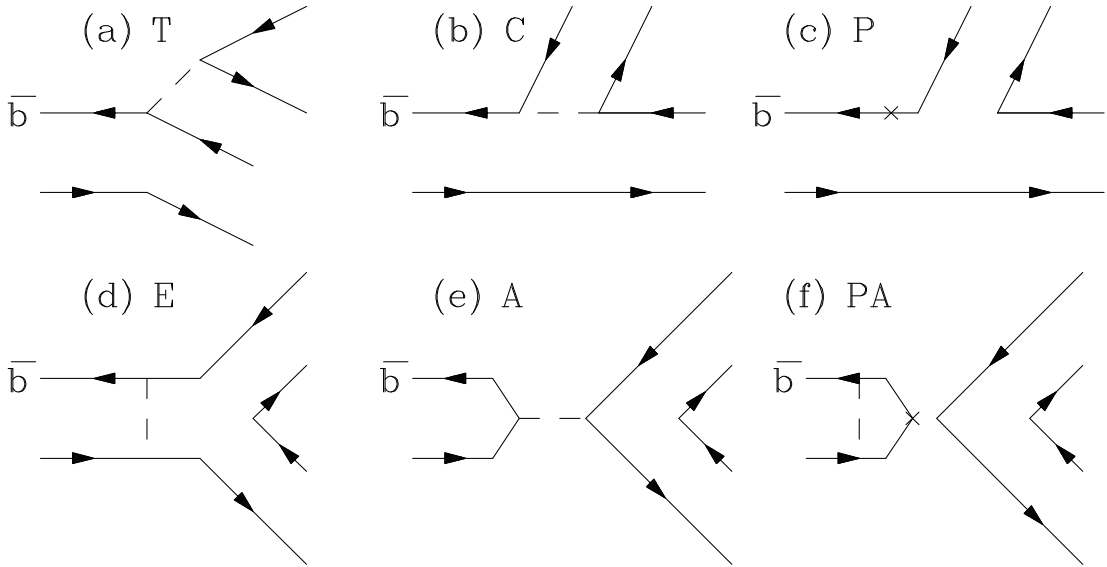}
\caption{Graphical representation of invariant amplitudes describing
$B$-meson decays.  (a) Color-favored tree ($T$); (b) Color-suppressed tree
($C$); (c) Penguin ($P$); (d) Exchange ($E$); (e) Annihilation ($A$); (f)
Penguin annihilation ($PA$).  Dashed lines indicate $W$ exchanges; $\times$
denotes a penguin $\bar b \to \bar d$ or $\bar b \to \bar s$ insertion.
\label{fig:amps}}
\end{figure}

The decays of $B$ mesons to two-body final states provide rich data for
determining parameters of the Cabibbo-Kobayashi-Maskawa (CKM) matrix
which is thought to describe the observed violations of CP symmetry.
These processes also yield valuable tests of the SU(3) flavor symmetry
obeyed by final-state $u$, $d$, and $s$ quarks.  Following early SU(3)
analyses of $B$ decays \cite{Zeppenfeld:1980ex,Savage:1989ub,Chau:1990ay},
a hierarchy of invariant amplitudes was established, based on a convenient
graphical language \cite{Gronau:1994rj}.  Dominant amplitudes were found to
be color-favored tree ($T$) followed by color-suppressed tree ($C$) and
penguin ($P$).  These three amplitudes involve only the decaying $\bar b$
quark in the initial $B$ meson and hence are 
approximately independent of the light
``spectator'' quark.  Amplitudes considerably suppressed in comparison
with them, all of which require participation of the spectator quark, are
exchange ($E$), annihilation ($A$), and penguin annihilation ($PA$).  
All six amplitudes are illustrated in Fig.\ \ref{fig:amps}. 

As pointed out a number of years ago \cite{Blok:1997yj,Blok:1997yw}, effects of
the amplitudes $E$, $A$, and $PA$ can also be generated by rescattering from
processes whose amplitudes [color-favored tree ($T$), color-suppressed tree
($C$), or penguin ($P$)] do not involve $f_B$.  Since then, both
electron-positron and hadron collisions have yielded a wealth of information
on many suppressed processes, such as new limits on the branching fraction
for $B^0 \to K^+ K^-$ \cite{Aaij:2012as,Duh:2012ie} and observation of the decays
$B_s \to \pi^+ \pi^-$ \cite{Aaij:2012as,Aaltonen:2011jv} and $B^+ \to D_s^+
\phi$ \cite{Aaij:2012zh}.  In the present paper we study such processes
systematically, identifying promising intermediate states contributing to
rescattering.  We find that the suppressed processes have typical $E$ 
amplitudes ranging from 5\% to 10\% of the largest amplitude contributing 
to rescattering, while $PA$ amplitudes are somewhat larger.
Based on this regularity, and using relations based on U-spin or on time-reversal,
we predict the branching fractions for a large number of as-yet-unseen $B$ 
and $B_s$ decays.

Calculations of $E, A$ and $PA$-type amplitudes in QCD factorization are quite challenging.
In $B$ decays with one charmed meson in the final state these amplitudes involve 
unknown matrix elements of non-local four-quark operators \cite{Mantry:2003uz}, while 
$E/A/PA$ amplitudes for charmless decays depend on divergent integrals \cite{Beneke:2003zv}.
Refs. \cite{Lu:2001yz,Xiao:2011tx,Chang:2012xv} and a few references quoted therein 
have presented model-dependent attempts to calculate $E, A$ and $PA$ amplitudes 
within QCD.  

In Section \ref{sec:AandE} we outline our strategy for evaluating rescattering
contributions to suppressed $E$, $A$, and $PA$ amplitudes.  In Section 
\ref{sec:relmag} we use current data to obtain ranges of ratios characterizing 
the suppression of these amplitudes relative to relevant $T, C$ and $P$ amplitudes.  
We then apply these ratios in Section \ref{sec:pred} to predict branching ratios for 
a number of $B$ and $B_s$ decays. Section \ref{sec:SU3} highlights predictions 
based on flavor SU(3) and time-reversal invariance, while Section 
\ref{sec:sum} concludes.

\section{
\boldmath
$E, A$ and $PA$ amplitudes from rescattering
\label{sec:AandE}}

\begin{figure}
\includegraphics[width=0.98\textwidth]{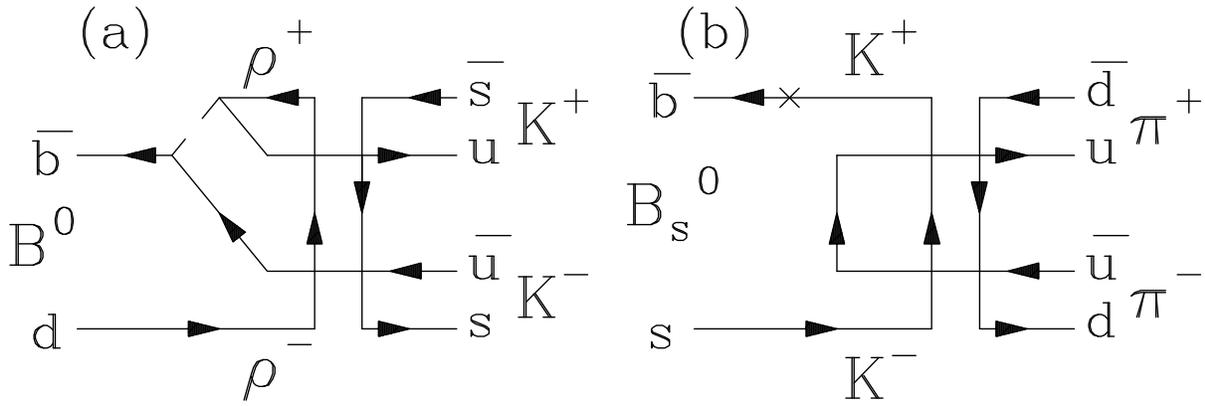}
\caption{Rescattering contributions. (a) To $B^0 \to K^+ K^-$; initial tree
($T$) amplitude, $\rho^+ \rho^-$ intermediate state contributing to exchange
($E$) amplitude.  (b) To $B_s \to \pi^+ \pi^-$; initial penguin ($P$)
amplitude, $K^+K^-$ intermediate state contributing to penguin annihilation
($PA$) amplitude.
\label{fig:epax}}
\end{figure}

\begin{figure}
\begin{center}
\includegraphics[width = 0.6\textwidth]{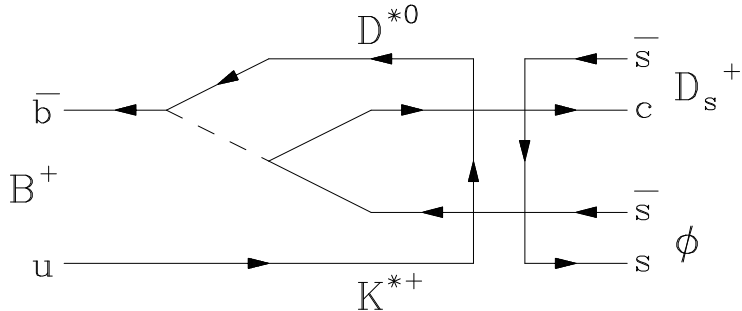}
\end{center}
\caption{Rescattering contributions to $B^+ \to D_s^+ \phi$ from a $D^{*0}
K^{*+}$ intermediate state whose amplitude is of the color-suppressed tree
($C$) form.
\label{fig:ax}}
\end{figure}
\begin{figure}
\begin{center}
\includegraphics[width = 0.6\textwidth]{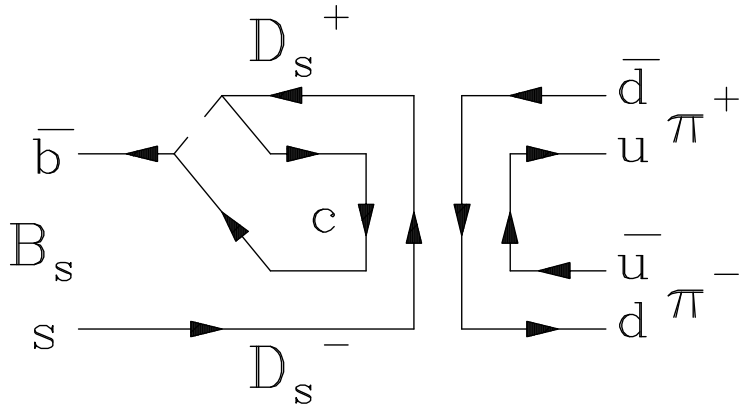}
\end{center}
\caption{Rescattering contributions to $B_s \to \pi^+ \pi^-$ from a $D_s^+ D_s^-$ 
intermediate state whose amplitude is of the color-favored tree ($T$) form.
\label{fig:double}}
\end{figure}
The manner in which a suppressed amplitude is generated by rescattering
can be illustrated by some examples.  Fig.\ \ref{fig:epax}\,(a) depicts the
contribution to an exchange ($E$) amplitude for $B^0 \to K^+ K^-$ from the
$\rho^+ \rho^-$ intermediate state, where the initial amplitude is of the tree
($T$) form. Fig.\ \ref{fig:epax}\,(b) describes a penguin annihilation ($PA$) 
amplitude for $B_s \to \pi^+ \pi^-$ obtaining a contribution from a $K^+ K^-$ 
intermediate state, where the initial amplitude is of the penguin ($P$) form.
Fig.\ \ref{fig:ax} shows the contribution of a $D^{*0} K^{*+}$ intermediate
state [initial amplitude of the color-suppressed tree ($C$) form] to an
annihilation ($A$) amplitude for $B^+ \to D_s^+ \phi$. 
Finally, Fig.\ \ref{fig:double} shows the contribution of a $D_s^+D_s^-$ 
intermediate state (from $T$) to a penguin annihilation ($PA$) amplitude
in $B_s \to \pi^+\pi^-$. 

In $B$ decays, whose average multiplicity is quite large, a given final state
can be generated by rescattering from any number of intermediate states, many
of which have not yet been observed.  Even if they were seen, it would not
be clear with what relative phases their contributions should be added
together.  We do expect (quasi) two-body intermediate states to dominate,
because rescattering from three-body or higher multiplicity states
to two-body final states is expected to be greatly suppressed.  For instance,
while momenta are fixed for decays to two particles, they
fill the plane of the Dalitz plot for three-body decays.
We assume that rescattering is dominated by light-quark exchange.
Rescattering due to heavy charm-quark exchange, depicted 
in Fig.\,\ref{fig:double}, is highly suppressed and will be mentioned 
briefly at the end of Section \ref{sec:relmag}.

In order to circumvent the shortcoming due to having several contributing
states,  we identify the (quasi) two-body
intermediate state with the largest branching fraction, whose $T$, $C$, or $P$
amplitude we compare with the $E$, $A$, or $PA$ amplitude of the suppressed
process.  For several such processes, we find that the ratio $|E/T|$ 
lies within a narrow range of values between 0.05 and 0.10 while $|PA/P|$ is 
between 0.15 and 0.20.  Finding no experimental evidence for a nonzero $|A/T|$,
we will assume that this ratio takes values in the same range as $|E/T|$.  
The values of these three ratios are then used to predict branching fractions 
for a large number of the suppressed processes originally identified in 
Ref.\ \cite{Blok:1997yj,Blok:1997yw}.

For simplicity we limit our consideration to intermediate states with two
pseudoscalar mesons ($PP$), one pseudoscalar and one vector ($PV$), and
two vector ($VV$) mesons.  The states contributing to $PP$ and $PV$ final
states are summarized in Table \ref{tab:ints}.

\begin{table}
\caption{$PP$, $PV$, and $VV$ intermediate states contributing to $B \to PP,PV$
decays.  Other states are forbidden to contribute by parity conservation in
the strong interactions.
\label{tab:ints}}
\begin{center}
\begin{tabular}{c c} \hline \hline
Final &     Contributing      \\
State & intermediate state(s) \\ \hline
 $PP$ & $PP$, $(VV)_{L=0,2}$  \\
 $PV$ & $PV$, $(VV)_{L=1}$    \\ \hline \hline
\end{tabular}
\end{center}
\end{table}

\section{Relative magnitude of suppressed amplitudes
\label{sec:relmag}}

We begin by reviewing the status of the 
suppressed decays discussed in Ref.\ \cite{Blok:1997yj,Blok:1997yw}.  
Table \ref{tab:BPP} lists the $PP$ decays of
nonstrange $B$ mesons with examples of contributing intermediate
states.  Tables \ref{tab:BsPP} 
lists the corresponding final and intermediate states for $B_s$ decays.
We note four isospin relations between $B_s$ decay amplitudes to charged 
and neutral mesons,
\bea\label{isospin}
A(B_s \to D^+D^-) & = & -A(B_s \to D^0 \bar D^0)~~~~~~~~~~(\Delta I =0)~, \nonumber\\
A(B_s \to \pi^+\pi^-) & = & -\s\,A(B_s \to \pi^0\pi^0)~~~~~~~(\Delta I = 0)~, \nonumber\\
A(B_s \to D^+ \pi^-) & = & -\s\,A(B_s \to D^0\pi^0)~~~~~~(\Delta I = 1/2)~, \nonumber\\
A(B_s \to D^-\pi^+) & = & -\s\,A(B_s \to \bar D^0 \pi^0)~~~~~~(\Delta I = 1/2)~.\label{pairsBs}
\eea 
%
\begin{table}
\caption{$E/A$-type decays of nonstrange $B$ mesons to two
pseudoscalars, and $T$-type decays to intermediate states contributing 
to these decays by rescattering.
Measured $E/A/PA$ decays (first line or two lines in each subtable), along
with possible contributing rescattering decays (subsequent lines).
The branching ratios for all measured decays are given, along with
the value of $R$ found for each rescattering decay, assuming that it
is dominant. The ratio of amplitudes probed in each $E/A/PA$ decay is
given at the top of the ``ratio'' column. For entries which are
flagged with letters [such as (a)] further details are given in the text.
\label{tab:BPP}}
\center
\begin{tabular}{cccccc} \hline \hline
 CKM &  Decay & Type & Int. State & BR & Ratio $R$ \\ \hline
$V^*_{cb} V_{ud}$ & $\bd \to D_s^- K^+$ & $E$ &   & $(2.31 \pm 0.24) \times
 10^{-5}$ \cite{HFAG} & $|E/T|$ \\
 & & & $D^-\pi^+$ & $(2.68 \pm 0.13)\times 10^{-3}$& $0.09 \pm 0.01$ \\
 & & & $D^{*-}\rho^+$ & $0.96\cdot(6.8 \pm 0.9)\times 10^{-3}$ (a)
 & $0.06 \pm 0.01$ \\ \hline
$V^*_{cb} V_{cd}$ & $\bd \to D^0 \overline {D}^0$ & $E$ & & $< 4.3 \times 10^{-5}$
 & $|E/T|$ \\
 & $\bd \to D_s^+ D_s^-$ & $E$ & & $< 3.6 \times 10^{-5}$ & $|E/T|$ \\
 & & & $D^+ D^-$ & $(2.11 \pm 0.31)\times 10^{-4}$ & $<0.4$ \\
 & & & $D^{*+} D^{*-}$ & (7.0$\pm$0.8)$\times$10$^{-4}$ (b)
 & $<0.2$ \\  \hline
$V^*_{ub} V_{ud}$ & $\bd \to K^+ K^-$ & $E$ & & $< 2 \times 10^{-7}$
 \cite{Duh:2012ie} & $|E/T|$ \\
 & & & $\pi^+\pi^-$ & $(5.15 \pm 0.22)\times 10^{-6}$ & $<0.2$  \\
 & & & $\rho^+\rho^-$ & $(2.42 \pm 0.31)\times 10^{-5}$ (c) & $<0.1$ \\ \hline
$V^*_{ub} V_{cs}$ & $B^+ \to D^+ K^0$ & $A$  & & $< 2.9 \times 10^{-6}$
 & $|A/T|$  \\
 & & & $D_s^+ \pi^0$ (d) & $(1.6 \pm 0.5)\times 10^{-5}$ & $<0.3$  \\ \hline
$V^*_{ub} V_{cd}$ & $\bd \to D_s^+ K^- $ & $E$ & & $-$ & $-$ \\ 
& & & $D^+\pi^-$ &  $(7.8 \pm 1.4)\times 10^{-7}$ & $-$ \\ \hline
$V^*_{ub} V_{cd}$ & $B^+\to D_s^+\overline {K}^0$ & $A$ & & $<8\times 10^{-4}$ & $-$\\
& & & $D^+\pi^0$ (d) & $-$ & $-$ \\ \hline \hline
\end{tabular}
\end{table}

One can also list a number of nonstrange $B$ decays through $E$ and $A$
amplitudes to $PV$ final states.  (No such $B_s$ decays have been reported
yet.)  These are given in Table \ref{tab:BVP} with examples of contributing
non-suppressed intermediate states.  All branching ratios quoted in Tables 
\ref{tab:BPP}, \ref{tab:BsPP} and  \ref{tab:BVP} are taken from the Particle 
Data Group \cite{pdg}, unless otherwise indicated. 
Finally, one can consider suppressed $B \to VV$ decays by replacing both 
pseudoscalars in Tables \ref{tab:BPP} and \ref{tab:BsPP} by vector mesons.

\begin{table}
\caption{$E/PA$-type decays of $B_s$ mesons to two pseudoscalars,
and $T$-type decays to intermediate states contributing to these decays by rescattering.
Information organized as in Table \ref{tab:BPP}.
\label{tab:BsPP}}
\center
\begin{tabular}{cccccc} \hline \hline
CKM & Decay & Type & Int. state & BR & Ratio $R$ \\ \hline
$V^*_{cb} V_{cs}$ & $B_s \to D^+ D^-$ & $E$ & & $-$ & $|E/T|$ \\
 & $B_s \to D^0 {\bar D}^0$ & $E$ & & $-$ &  $|E/T|$ \\
 & & & $D_s^+ D_s^-$ & $(5.3 \pm 0.9)\times 10^{-3}$ & $-$ \\
 & & & $D_s^{*+} D_s^{*-}$ & (1.60$\pm$0.29)$\times$10$^{-2}$
 (e) & $-$  \\ \hline
 $V^*_{cb} V_{cs}$ & $B_s \to \pi^+ \pi^-$ & $PA$ & & $(0.73 \pm 0.14)\times 10^{-6}$ 
 \cite{HFAG} & $|PA/P|$ \\
 & $B_s \to \pi^0 \pi^0$ & $PA$/$\sqrt{2}$ & & $-$ & $|PA/P|$   \\
  & & & $K^+ K^-$ & $(2.45 \pm 0.18)\times 10^{-5}$ \cite{HFAG} & $0.17 \pm 0.02$\\
    & & & $K^{*0} \bar K^{*0}$ & $(1.7 \pm 0.5)\times 10^{-5}$ (f) 
    & $0.21 \pm 0.04$ \\    \hline
$V^*_{cb} V_{us}$ & $B_s \to D^- \pi^+$ & $E$ & & $-$ & $|E/T|$ \\
 & $B_s \to  {\bar D}^0 \pi^0$ & $E$/$\sqrt{2}$ & & $-$ & $|E/T|$ \\
 & & & $D_s^- K^+$ & (g) & $-$ \\ \hline
$V^*_{ub} V_{cs}$ & $B_s \to D^+ \pi^-$ & $E$ & & $-$ & $|E/T|$ \\
 & $B_s \to D^0 \pi^0$ & $E/\sqrt{2}$ & & $-$ & $|E/T|$ \\  
 & & & $D_s^+ K^-$ & (g) & $-$ \\  \hline \hline
  $V^*_{cb} V_{cs}$ & $B_s \to \pi^+ \pi^-$ & $PA$ & & $(0.73 \pm 0.14)\times 10^{-6}$ 
 \cite{HFAG} & $|PA/T|$ \\
 & $B_s \to \pi^0 \pi^0$ & $PA$/$\sqrt{2}$ & & $-$ & $|PA/T|$   \\
 & & & $D_s^+ D_s^-$ & $(5.3 \pm 0.9)\times 10^{-3}$ & $0.012 \pm 0.002$ \\
& & & $D_s^{*+} D_s^{*-}$ & (1.60$\pm$0.29)$\times$10$^{-2}$
 &  $0.007 \pm 0.002$ \\
\hline \hline
\end{tabular}
\end{table}

As noted in the introduction, rescattering can occur via many intermediate
states.  We can identify at most a few of them, but there will always be one
with the largest branching fraction.  We can use that one to calculate a
``typical'' ratio of the suppressed amplitude to the largest unsuppressed one.
We then have to assume that the effect of many intermediate states (whether
constructive, incoherent, or destructive) is roughly the same for all cases. 
With this in mind, we calculate the amplitude ratio for all measured
$E/A/PA$-type decays, assuming a single intermediate state, that with the
largest branching fraction. 

\begin{table}
\caption{E/A-type decays of nonstrange $B$ mesons to $PV$ final states,
and $T$ or $C$-type decays to intermediate states contributing to these decays
by rescattering. Information organized as in Table \ref{tab:BPP}.
\label{tab:BVP}}
\center
\begin{tabular}{cccccc} \hline \hline
 CKM  & Decay & Type  & Int. State & BR & Ratio $R$ \\
\hline
$V^*_{cb} V_{ud}$ & $\bd \to D_s^{*-} K^+$ & $E$ & & $(2.19 \pm 0.30)
 \times 10^{-5}$ & $|E/T|$\\
 & $\bd \to D_s^- K^{*+}$ & $E$ &  & $(3.5 \pm 1.0)\times 10^{-5}$ & $|E/T|$ \\
 & & & $D^{*-}\pi^+$ & $(2.76 \pm 0.13)\times 10^{-3}$ & $0.09 \pm 0.01$ \\
 & & & $D^- \rho^+$ & $(7.8 \pm 1.3)\times 10^{-3}$ & $0.05 \pm 0.01$ \\
  & & & $D^{*-} \rho^+$ & (2.7$\pm$0.4)$\times$10$^{-4}$ (h) & \\
 \hline
$V^*_{cb} V_{cd}$ & $\bd \to D^{*0} \overline{D}^0, D^0 \overline {D}^{*0}$
 & $E$ & & $< 2.9 \times 10^{-4}$ & $|E/T|$\\
 & $\bd \to D_s^\pm D_s^{*\mp}$ & $E$ & &  $< 1.3 \times 10^{-4}$ & $|E/T|$ \\
& & & $D^{*+} D^-$ & $(6.1 \pm 1.5)\times 10^{-4}$ & $<0.5$ \\
\hline
$V^*_{ub} V_{ud}$ & $\bd \to K^{*\pm} K^\mp$ & $E$ & & $-$ & $|E/T|$ \\
 & & & $\rho^{\pm}\pi^{\mp}$ & $(2.30 \pm 0.23)\times 10 ^{-5}$ & $-$\\
 \hline
$V^*_{ub} V_{cs}$  & $B^+ \to D^+ K^{*0}$ & $A$ & &  $< 1.8 \times 10^{-6}$ \cite{Aaij:2012zh}
& $|A/T(C)|$\\
& $B^+ \to D^{*+} K^0$ & $A$ & & $<9.0\times 10^{-6}$ & $|A/T(C)|$ \\
& & & $D_s^{*+} \pi^0~(T)$ & $<2.6\times 10^{-4}$ & $-$ \\
& & & $D_s^+ \rho^0~~(T)$ & $< 3.0\times 10^{-4}$ & $-$\\ 
& & & $D^0 K^{*+}(C)$ & $\sim 1\times 10^{-5}$ (i,j)& $<0.4$\\ 
\hline
$V^*_{ub}V_{cs}$  & $B^+ \to D_s^+ \phi$ & $A$ &  & $(1.87^{+1.30}_{-0.82}) \times 10^{-6}$
\cite{Aaij:2012zh} & $|A/T(C)|$ \\
& & & $D^0 K^{*+}(C)$ & $\sim 1\times 10^{-5}$ (i,j)
 & $\sim 0.4 \pm 0.1$ \\
& & & $D^+_s \omega~(T)$ & $<4\times 10^{-4}$ (k) & (l) \\ \hline
$V^*_{ub} V_{cd}$ & $\bd \to K^{*-} D_s^+$ & $E$ & & $-$ & $|E/T|$ \\
& $\bd \to K^- D_s^{*+}$ & $E$ &  & $-$ & $|E/T|$\\
& & & $D^+ \rho^-$ & $-$ & $-$ \\
& & & $D^{*+} \pi^-$ & $-$ & $-$ \\ 
\hline
$V^*_{ub} V_{cd}$ & $B^+ \to \overline{K}^{*0} D_s^+$ & $A$ &
 & $<4.4\times 10^{-6}$ \cite{Aaij:2012zh} & $|A/T|$ \\
& $B^+ \to \overline{K}^0 D_s^{*+}$ & $A$ & & $-$ & $|A/T|$ \\
& & & $D^+ \rho^0$ & $-$ & $-$ \\
& & & $D^{*+} \pi^0$ & $-$ & $-$ \\
\hline \hline
\end{tabular}
\end{table}

Amplitudes are evaluated as square roots of branching fractions,
with phase-space differences ignored.  
We consider only suppressed amplitudes and amplitudes for intermediate states 
which share the same CKM factor. Thus, for instance, in the amplitude $E$ for 
$B^0 \to K^+ K^-$ involving $V^*_{ub}V_{ud}$ we ignore a rescattering 
contribution  from $B^0 \to D^+ D^-$ involving $V^*_{cb}V_{cd}$. 
For the amplitude ratio (i.e., the rescattering suppression factor), for which 
CKM factors associated with the decays in the numerator and denominator cancel,
we use the symbol $R \equiv |[E/A/PA]/[T/C/P]|$.  

For $B\to PP$ $E/A/PA$-type decays, we give the $PP$ and $VV$ intermediate states;
for $B\to PV/VP$ $E/A/PA$ decays, we give the $PV$, $VP$ and $VV$ intermediate
states. Now, for the $VV$ intermediate states, Ref.\ \cite{pdg}  gives the
branching ratio for the decay to all three helicity states. However, only two
(one) of these -- those with positive (negative) parity -- contribute to
rescattering to $PP$ ($PV/VP$) final states.  Thus, the effective rescattering
branching ratio is probably smaller than that given in the tables, and the
value of $R$ larger.  For many $B\to VV$ decays, the polarization fractions
have been measured. This allows us to modify the total branching ratios
appropriately, which we do where possible.

In Tables \ref{tab:BPP}, \ref{tab:BsPP} and \ref{tab:BVP} we list all 
measured $E/A/PA$-type decays, along with the value of $R$ obtained from 
individual decays into intermediate rescattering states.  Some of the quoted 
branching ratios of the latter processes require some details which we give now.
\medskip

\noindent
(a) The helicity amplitudes for $\bd \to D^{*-} \rho^+$ were measured in
Ref.~\cite{CLEOD*rho}, with the result $|H_0| = 0.941$, $|H_\|| = 0.27$,
$|H_\perp| = 0.21$.  The fraction of decays with positive parity is thus
$f_+ = (|H_0|^2 + |H_\||^2)/(|H_0|^2 + |H_\||^2 + |H_\perp|^2) = 0.96$. This
indicates that the rescattering of $\bd \to D^{*-} \rho^+$ contributes
significantly to $\bd \to D_s^- K^+$. On the other hand, the fraction of decays
with negative parity is 0.04, so that there is little rescattering contribution
to $\bd \to D_s^- K^{*+}$. 
\medskip

\noindent
(b) The fraction of $B^0 \to D^{*+}D^{*-}$ decays with positive parity is 
$f_+ = 0.850 \pm 0.025$ \cite{pdg}. We quote $f_+\br(B^0 \to D^{*+}D^{*-})
=(0.850 \pm 0.025)(8.2 \pm 0.9)\times 10^{-3} = (7.0 \pm 0.8)\times10^{-3}$.
\medskip

\noindent
(c) $B^0 \to \rho^+ \rho^-$ is dominated by longitudinal polarizations, 
$f_L = |H_0|^2/(|H_0|^2 + |H_\||^2 + |H_\perp|^2) = 
0.977^{+0.028}_{-0.024}$ \cite{pdg}.
\medskip

\noindent
(d) Decay amplitude is given by $T/\s$.
\medskip

\noindent
(e) We are assuming that the fraction of $B_s \to D_s^{*+} D_s^{*-}$ decays with
positive parity is the same as in $B^0 \to D^{*+}D^{*-}$.  This assumption is
supported by a calculation based on the heavy-quark expansion and factorization
\cite{Rosner:1990xx}. We quote $f_+\br(B_s \to D_s^{*+} D_s^{*-}) = 
(0.850 \pm 0.025)(1.88 \pm 0.34)\times 10^{-2} = (1.60 \pm 0.29)\times 10^{-2}$.

\medskip

\noindent
(f) The helicity amplitudes for $B_s \to K^{*0} {\bar K}^{*0}$ were
measured in Ref.~\cite{LHCbK*K*}, leading to $f_+ = 0.62 \pm 0.12$.
We quote $f_+\,\br(B_s \to K^{*0} \bar K^{*0})=(0.62 \pm 0.12)
(2.8 \pm 0.7)\times 10^{-5} = (1.7 \pm 0.5) \times 10^{-5}$,
$f_-\,\br( B_s \to K^{*0} \bar K^{*0}) = (1.1 \pm 0.4)\times 10^{-5}$.
\medskip

\noindent
(g) Using untagged $B_s$ decays only the charge-averaged branching ratio
has been measured, $\br(B_s \to D_s^{\pm}K^{\mp})=(2.9 \pm 0.6)\times 10^{-4}$
 \cite{pdg}.
\medskip

\noindent
(h) We quote $f_-\,\br(B^0 \to D^{*-}\rho^+) = 0.04\cdot (6.8 \pm 0.9)\times 
10^{-3} = (2.7 \pm 0. 4)\times 10^{-4}$.
\medskip

\noindent
(i) The decays $B^+ \to D^{(*)0} K^{(*)+}$ have not been measured, but the
decays $B^+ \to {\bar D}^{(*)0} K^{(*)+}$ have: $B(B^+ \to {\bar D}^0 K^{*+})
= (5.3 \pm 0.4) \times 10^{-4}$, $B(B^+ \to {\bar D}^{*0} K^+)=(4.20 \pm 0.34)
\times 10^{-4}$, $B(B^+ \to {\bar D}^{*0} K^{*+}) = (8.1 \pm 1.4) \times
10^{-4}$. BaBar has found that $r_B \equiv |A(B^+ \to D^0 K^{*+})|/|A(B^+ \to
{\bar D}^0 K^{*+})| = 0.31 \pm 0.07$ \cite{BaBarrB}.  This gives $B(B^+ \to
D^0 K^{*+}) = (0.31 \pm 0.07)^2 \cdot (5.3 \pm 0.4) \times 10^{-4} = (5.1 \pm
2.3) \times 10^{-5}$.
\medskip

\noindent
(j)  The isospin triangle relation, $A(B^0 \to D^0 K^{*0}) = A(B^+ \to D^0 K^{*+}) 
+ A(B^+ \to D^+ K^{*0})$, shown in Ref.~\cite{GGSSZ} implies that $r_B$ is 
smaller by at least  one $\sigma$ than its above-mentioned central value.  
With the experimental limits \cite{pdg}
$B(B^0 \to D^0 K^{*0}) < 1.1 \times 10^{-5}$ and 
$B(B^+ \to D^+ K^{*0}) < 1.8\times 10^{-6}$ \cite{Aaij:2012zh}, 
we have (in units of $10^{-3}$) $|A(B^0 \to D^0 K^{*0})| <3.3$ and 
$|A(B^+ \to D^+ K^{*0})| < 1.3$.  But taking $B(B^+ \to D^0 K^{*+}) =
(5.1 \pm 2.3) \times 10^{-5}$ yields $|A(B^+ \to D^0 K^{*+})| = 7.1 \pm 1.6$;
for the central value of this last branching ratio, the triangle does not
close.  It closes only if the branching ratio for $B^+ \to D^0 K^{*+}$
is at least $1.5\sigma$ below its central value. 
This, and independent supporting evidence discussed in the next point 
below, suggest that a likely value of $\br(B^+ \to D^0 K^{*+})$ is around 
$1\times 10^{-5}$, corresponding to  $r_B \simeq 0.15$.  
This value is consistent with a value $r_B= 0.115 \pm 0.045$  
obtained in Ref.\,\cite{CKMfitter} by a global fit to CKM parameters.
\medskip
 
\noindent
(k)  A potential rescattering state contributing to $B^+ \to D_s^+ \phi$ is 
$D_s^+ \omega$, for which one has a rather old upper bound 
$\br(B^+ \to D_s^+ \omega) <  4\times 10^{-4}$ \cite{Alexander:1993gp}.
An order of magnitude stronger upper bound, $\br(B^+\to D_s^+\omega) 
\lsim 1.2\times 10^{-5}$, is obtained  if one assumes 
$\br(B^+ \to D_s^+ \omega) \simeq \br(B^+ \to D_s^+ \rho^0)$ 
using an isospin relation, $\br(B^+ \to D_s^+ \rho^0) = \br(B^0 \to D_s^+ \rho^-)/2
< 1.2 \times 10^{-5}$ \cite{pdg}. We note that while $B^+ \to D^0 K^{*+}$ is due to a 
color-suppressed amplitude $C$, $B^+ \to D_s^+ \omega$ involves a 
color-favored tree amplitude $T/\s$ which is usually expected to be larger 
than $C$. Recalling our discussion of $B^+\to D^0 K^{*+}$ in point (j) we 
are led to conclude that both  $\br(B^+ \to D^0 K^{*+})$ and 
$\br(B^+ \to D_s^+ \omega)$ are most likely around $1\times 10^{-5}$.
Improved measurements of $\br(B^+ \to D_s^+ \omega)$ and 
$\br(B^+ \to D_s^+ \rho^0)$ (a potential dominant rescattering contributor to $B^+ \to 
D^+ K^{*0}$ and $B^+ \to D^{*+} K^0$), using the  BaBar, 
Belle and LHCb high statistics data, are of great importance. 
\medskip

\noindent
(l) The rescattering contribution of $B^+ \to D_s^+ \omega$ to $B^+ \to D_s^+
\phi$ due to $\omega$--$\phi$ mixing is OZI-suppressed \cite{OZI}. It is given
by $\br(B^+ \to D_s^+ \phi)_{\omega-\phi} = \br(B^+ \to D_s^+ \omega)\,
\delta^2$.  Here $\delta$ is the $\omega$--$\phi$ mixing 
angle, $\delta = -3.34^\circ$ or $\delta(m=m_\phi) = -4.64^\circ$ in
mass-independent or mass-dependent analyses \cite{Gronau:2009mp}. 
Assuming $\br(B^+ \to D_s^+ \omega) \sim 1\times 10^{-5}$ as argued above 
and taking a mass-dependent $\delta$, one finds 
$\br(B^+ \to D_s^+ \phi)_{\omega-\phi} \sim 0.7\times 10^{-7}$. This is only 
a tiny fraction of the measured value of $\br(B^+ \to D_s^+ \phi)$.

The information on ratios $R$ given in the last columns of Tables \ref{tab:BPP}, 
\ref{tab:BsPP} and \ref{tab:BVP} can be summarized as follows:
\begin{itemize}
\item The ratio $|E/T|$, obtained from $\br(B^0 \to D_s^- K^+)$, $\br(B^0\to
D^-\pi^+)$ and all their $VP$ analogues, lies in the narrow range $|E/T| = 0.05
- 0. 1$. This range describes well contributions of rescattering in $B^0 \to
D^-\pi^+, D^{*-} \rho^+ \to D^-_s K^+$ and $B^0 \to D^{*-} \pi^+, D^- \rho^+
\to D_s^{*-} K^+, D^-_s K^{*+}$. The different angular momenta involved in these
decays do not seem to affect much the value of $R$.  A number of other decay
modes involving $(d \bar d) \to (s \bar s)$ rescattering are expected to have
values of $R$ in the same range.  \item The ratio $|A/T|$ cannot be extracted
from $\br(B^+ \to D_s^+\phi)$ and $\br(B^+ \to D_s^+ \omega)$ because
rescattering from $D_s^+ \omega$ to $D_s^+\phi$ is OZI-suppressed. This seems
like a singular case, in which we are unable
 to identify a dominant intermediate state contributing to rescattering. 
 A less likely interpretation for the branching ratio of $B^+ \to D_s^+ \phi$ 
 is that physics beyond the CKM framework is at work.
\item The value of $|PA/P|$, obtained from $\br(B_s \to \pi^+\pi^-)$ and 
$\br(B_s \to K^+ K^-)$, is near 0.2, about twice the value of $|E/T|$.
In the last subtable in Table \ref{tab:BsPP} we also obtain a value for a ratio
$|PA/T|$, where $T$ is a color-favored tree amplitude determined by 
$\br(B_s \to D_s^{(*)+}D_s^{(*)-})$. This very small ratio of order 0.01,
corresponding to $D_s^{(*)+}D_s^{(*)-} \to \pi^+ \pi^-$ rescattering, is
suppressed by requiring two quark-antiquark rescatterings as shown in Fig.\
\ref{fig:double}.
Some portion of the suppression may be due to the exchange of the heavy charm
quark in rescattering.
 \end{itemize} 

\section{
\boldmath
Predictions based on ranges of $R$
\label{sec:pred}}

With the above-mentioned ranges of $R$ we can now predict the branching ratios
for other $E/A/PA$ decays. We will use the value $|E/T| = 0.07 \pm 0.02$ and
will {\em assume} the same range for $|A/T|$ in cases where one may identify a
potentially dominant $T$-type decay contributing by rescattering to an $A$-type
decay.  Finally, the value $|PA/P| = 0.17 \pm 0.02$ extracted from $\br(B_s
\to \pi^+\pi^-)$ and $\br(B_s \to K^+K^-)$ will be used to predict branching
ratios for $B_s$ decays into other pairs of unflavored mesons. The central 
values and uncertainties in the three ratios are chosen to describe  
ranges for these parameters. Thus the errors in predicted branching ratios, obtained by 
adding in quadrature these uncertainties and experimental errors in branching ratios,  
are not statistical. Rather, under our assumptions, they give reasonable ranges 
for a large number of branching ratios of decay modes which have not yet been observed.

Using the above values for the ratios $|E/T|$, $|A/T|$ and $|PA/P|$,
we obtain predictions for $B$ and $B_s$ decay branching ratios.
Results for $B, B_s \to PP$ and $B, B_s \to VP$ are presented in
Tables \ref{tab:predPP} and \ref{tab:predVP}, respectively.
Predictions appear in the first one or two lines in each subtable,
while the last line in each subtable quotes the corresponding largest
measured branching ratio for a process of type $T$ or $P$.  Entries in
the last subtable of Table \ref{tab:predPP} and in all but the second
subtable in Table \ref{tab:predVP} refer to CP-averaged branching
ratios which are measured using untagged $B^0$ and $B_s$ decays. Our
prediction for $\br(B^+ \to D^+ K^0)$ in Table \ref{tab:predPP} can
test our assumption $|A/T| = 0.07 \pm 0.02$.
%
\begin{table}[h]
\caption{Predictions for branching ratios of $B$ and $B_s$ decays to 
two pseudoscalar mesons. $E/A/PA$ decays appear in the first line or 
two lines in each subtable, while corresponding 
rescattering decay with largest branching ratio is given in the last line
of each subtable. 
Entries in the last subtable refer to CP-averaged branching ratios.
\label{tab:predPP}}
\center
\begin{tabular}{cccc} \hline \hline
CKM factor & Decay & Measured BR & Predicted BR \\ \hline
$V^*_{cb} V_{cd}$ & $B^0 \to D^0 {\bar D}^0$ & $< 4.3 \times 10^{-5}$ &
$(3.4 \pm 2.0)\times 10^{-6}$ \\
 & $B^0 \to D_s^+ D_s^-$ & $< 3.6 \times 10^{-5}$ & $(3.4 \pm 2.0)\times 10^{-6}$ \\
& $B^0 \to D^{*+} D^{*-}$ & $(7.0  \pm 0.8)\times 10^{-4}$ & \\ \hline
$V^*_{ub} V_{ud}$ & $B^0 \to K^+ K^-$ & $<2 \times 10^{-7}$ & 
$(1.2 \pm 0.7)\times 10^{-7}$ \\
& $B^0 \to \rho^+ \rho^-$ & $(2.42 \pm 0.31) \times 10^{-5}$ & \\ \hline
$V^*_{ub}V_{cd}$ & $B^0 \to D_s^+ K^-$ & $-$ & $(3.8  \pm 2.3)\times 10^{-9}$ \\
 & $B^0 \to D^+ \pi^-$ & $(7.8 \pm 1.4)\times 10^{-7}$ \\ \hline
$V^*_{ub} V_{cs}$ & $B^+ \to D^+ K^0$ & $< 2.9 \times 10^{-6}$ &  
$(1.6 \pm 1.0)\times 10^{-7}$ \\
& $B^+ \to D_s^+ \pi^0$ & $(1.6 \pm 0.5) \times 10^{-5}$  & \\ \hline
$V^*_{cb} V_{cs}$ & $B_s \to D^+ D^-$ & $-$ & $(7.8 \pm 4.7)\times 10^{-5}$ \\
& $B_s \to D^0 \bar D^0$ &$-$ & $(7.8 \pm 4.7)\times 10^{-5}$ \\
& $B_s \to D_s^{*+}D_s^{*-}$ & $(1.60 \pm 0.29)\times 10^{-2}$ &  
 \\ \hline
$V^*_{cb} V_{us}, V^*_{ub}V_{cs}$ & $B_s \to D^{\pm}\pi^{\mp}$ & $-$ & 
$(1.4 \pm 0.9)\times 10^{-6}$ \\
& $B_s \to D^0 \pi^0, \bar D^0 \pi^0$ & $-$ & $(0.7 \pm 0.4)\times 10^{-6}$ \\
& $B_s \to D_s^{\pm} K^{\mp}$ & $(2.9 \pm 0.6)\times 10^{-4}$ & \\ \hline \hline
\end{tabular}
\end{table}
\begin{table}[h]
\caption{Predictions for branching ratios of $B$ and $B_s$ decays to vector
and pseudoscalar mesons organized as in Table \ref{tab:predPP}.
Entries in all but the second subtable  refer to CP-averaged branching ratios.
\label{tab:predVP}}
\center
\begin{tabular}{cccc} \hline \hline
CKM factor & Decay & Measured BR & Predicted BR \\ \hline
$V^*_{cb}V_{cs}$ & $B_s \to D^{*\pm} D^{\mp}$ & $-$  & $(6.1 \pm 3.6)\times 10^{-5}$ \\
& $B_s \to D^{*0} \bar D^0, D^0 \bar D^{*0}$ & $-$ & $(6.1 \pm 3.6)\times 10^{-5}$ \\
& $B_s \to D_s^{*\pm} D_s^{\mp}$ & $(1.24 \pm 0.21)\times 10^{-2}$ & \\ \hline
$V^*_{cb}V_{cs}$ & $B_s \to \rho^+ \pi^-$ & $-$ & $(3.1 \pm 1.4)\times 10^{-7}$ \\
& $B_s \to \rho^- \pi^+$ & $-$ & $(3.1 \pm 1.4)\times 10^{-7}$ \\
& $B_s \to K^{*0} \bar K^{*0}$ & $(1.1 \pm 0.4)\times 10^{-5}$ (f) & \\ \hline
$V^*_{cb} V_{cd}$ & $B^0 \to D_s^{*\pm} D_s^{\mp}$ & 
$< 1.3 \times 10^{-4}$ & $(3.0 \pm 1.9)\times 10^{-6}$ \\
 & $B^0 \to D^{*0} \bar D^0, D^0 \bar D^{*0}$ &
  $<2.9\times 10^{-4}$ & $(3.0 \pm 1.9)\times 10^{-6}$ \\
& $B^0 \to D^{*\pm} D^\mp$ & $(6.1 \pm 1.5) \times 10^{-4}$ & \\  \hline
$V^*_{ub} V_{ud}$ & $B^0  \to K^{*\pm} K^\mp$ & $-$ & $(1.1 \pm 0.7)\times 10^{-7}$\\
& $B^0 \to \rho^{\pm} \pi^{\mp}$ & $(2.30 \pm 0.23)\times 10^{-5}$ & \\ \hline \hline
\end{tabular}
\end{table}

\section{Predictions based on flavor SU(3) or time-reversal 
\label{sec:SU3}}
In Eqs.\,(\ref{isospin}) we have presented four isospin relations in pairs of $E$ 
and $PA$-type $B_s$ decay amplitudes, leading to relations between
corresponding decay branching ratios. Other relations among 
$E$ and $A$-type $B$ and $B_s$ decay amplitudes follow in the limit of flavor SU(3) 
symmetry. Two subgroups of SU(3), U-spin and V-spin of which $(d, s)$ and $(u, s)$
are fundamental doublet representations, are useful in deriving these relations.
We will focus our attention on relations for decays into two pseudoscalar mesons,
discussing in certain cases also relations for $B, B_s \to VP$ and $B, B_s \to VV$. 

In the V-spin symmetry limit, applying $u \leftrightarrow s$ reflection, 
one has
\beq
A(B^0 \to D_s^+ D_s^-) = A(B^0 \to D^0 \bar D^0)~,
\eeq  
as assumed in Tables \ref{tab:BPP} and \ref{tab:predPP}.
Thus, in the V-spin symmetry limit the two corresponding branching ratios
are predicted to be equal.

Using approximate symmetry of strong interactions under U-spin reflection, 
$d \leftrightarrow s$, and considering the U-spin structure of the effective weak 
Hamiltonian and of initial and final states, we find:
\bea\label{BsBPP}
A(B_s \to D^- \pi^+) & = & \lambda\,A(B^0 \to D_s^- K^+)~,\nonumber \\
-\lambda\,A(B_s \to D^+ D^-) & = & A(B^0 \to D^+_s D^-_s)~,\nonumber \\
-\lambda\,A(B_s \to D^+ \pi^-)& = & A(B^0 \to D^+_s K^-) ~,
\eea
and 
\beq
A(B^+ \to D_s^+ \bar K^0) = -\lambda\,A(B^+\to D^+ K^0)~.
\eeq
Here  $\lambda \equiv V_{us}/V_{ud} \approx -V_{cd}/V_{cs} = 0.231$ \cite{pdg}.
Given the value of $\br(B^0 \to D^-_s K^+)$  
in Table \ref{tab:BPP}, the first of Eqs.\,(\ref{BsBPP}) leads to  predicting
$\br(B_s \to D^-\pi^+)$ in the U-spin symmetry limit,
\beq
\br(B_s \to D^- \pi^+)  \simeq  (1.23 \pm 0.13)\times 10^{-6}~.
\eeq
This value is in agreement with the prediction for the CP-averaged branching ratio 
quoted in Table \ref{tab:predPP} which involves a larger uncertainty.
In a similar manner one has U-spin relations for corresponding $B, B_s \to VP$ 
decays, such as
\bea\label{BsBVP}
A(B_s \to D^{*-} \pi^+) & = & \lambda\,A(B^0 \to D_s^{*-} K^+)~,\nonumber \\
A(B_s \to D^- \rho^+) & = & \lambda\,A(B^0 \to D_s^- K^{*+})~.
\eea
Taking branching ratios quoted in Table \ref{tab:BVP}, we obtain
\bea
\br(B_s \to D^{*-} \pi^+) & = & (1.2 \pm 0.2)\times 10^{-6}~,\nonumber \\
\br(B_s \to D^- \rho^+) & = & (1.9 \pm ~0.5)\times 10^{-6}~.
\eea
These predictions add to those already given in Table \ref{tab:predVP}.

We will now show that the predictions obtained in Section \ref{sec:pred}, 
assuming a dominant rescattering contribution in $E$-type decays, 
are consistent with U-spin relations such as Eqs.\,(\ref{BsBPP}) and (\ref{BsBVP}). 
We will use the fact that final states on the left-hand side of these equations 
are rescattering states contributing to corresponding amplitudes on the 
right-hand side, while final states on the right-hand side contribute as rescattering 
states to amplitudes on the left-hand side.  

Let us focus, for instance, on the first U-spin relation in Eqs.\,(\ref{BsBPP})
between two $E$-type amplitudes. We will show now that this relation may be derived 
using our assumption of dominant rescattering states for which two respective 
$T$-type amplitudes are related to each other by U-spin. We are assuming that 
$B^0 \to D^-_s K^+$ is dominated by a positive-parity $D^{*-} \rho^+$
rescattering state, 
\beq\label{effective1}
|A(B^0 \to D^-_s K^+)| =  |A(B^0 \to [D^{*-}\rho^+]_+)|\,|A([D^{*-}\rho^+]_+ \to D_s^- K^+)|~,
\eeq
where 
\beq\label{+}
|A(B^0 \to [D^{*-}\rho^+]_+)| \equiv \sqrt {|A(B^0 \to [D^{*-}\rho^+]_0)|^2   + 
|A(B^0 \to [D^{*-}\rho^+]_\parallel)|^2}~.
\eeq
Similarly one obtains
\beq\label{effective2}
|A(B_s \to D^- \pi^+)| = |A(B_s \to [D_s^{*-} K^{*+}]_+)|\,|A([D_s^{*-} K^{*+}]_+\to D^- \pi^+)|~,
\eeq 
where dominance of $B_s \to [D_s^{*-} K^{*+}]_+$ over  $B_s \to D_s^- K^+$ is 
implied by $|A(B^0 \to [D^{*-} \rho^+]_+)|$ $ > |A(B^0 \to D^- \pi^+)|$ and U-spin symmetry.  

Assuming that the rescattering amplitude is invariant under U-spin, 
$A([D_s^{*-} K^{*+}]_+\to D^- \pi^+) = A([D^{*-}\rho^+]_+ \to D_s^- K^+)$, we obtain
the first of Eqs.\,(\ref{BsBPP}) as required,
\beq\label{ratio-lambda}
\frac{|A(B_s \to D^- \pi^+)|}{|A(B^0 \to D^-_s K^+)|} = 
\frac{|A(B_s \to [D_s^{*-} K^{*+}]_+)|}{|A(B^0 \to [D^{*-}\rho^+]_+)|} = \lambda~.
\eeq
The second equality, giving the ratio of two positive parity $T$-type amplitudes, 
follows from the behavior under  U-spin reflection of the effective weak 
Hamiltonian and of initial and final states.  

At this point we wish to comment on the definition of the magnitude of the effective rescattering 
amplitude for positive parity, $|A([D^{*-}\rho^+]_+ \to D_s^- K^+)|$ in (\ref{effective1}),
which we have defined in Table \ref{tab:BPP} as the ratio $|E/T|=0.06 \pm 0.01$ in 
$B^0 \to D_s^- K^+$.
Eq.\,(\ref{effective1}) may be expanded,
\bea\label{B0DsK}
A(B^0 \to D^-_s K^+) & = & A(B^0 \to [D^{*-}\rho^+]_0)\,A([D^{*-}\rho^+]_0 \to D_s^- K^+)
\nonumber\\
& + & A(B^0 \to [D^{*-}\rho^+]_\parallel)\,A([D^{*-}\rho^+]_\parallel \to D_s^- K^+)~,
\eea 
where $[D^{*-}\rho^+]_{0,\parallel}$ are longitudinal and parallel polarization states, and 
$A([D^{*-}\rho^+]_{0,\parallel} \to D_s^- K^+)$ are corresponding strong interaction rescattering 
amplitudes. The $B^0 \to D_s^- K^+$ decay rate is obtained by squaring the above sum and
integrating over the angular dependence of the two pairs of final pseudoscalars, 
$\bar D^0 \pi^-$ (or $D^- \pi^0$) and $\pi^+\pi^0$.  The interference term drops by integration 
implying (we omit phase-space factors),
\bea
|A(B^0 \to D^-_s K^+)|^2 & = & |A(B^0 \to [D^{*-}\rho^+]_0)|^2\,|A([D^{*-}\rho^+]_0 \to D_s^- K^+)|^2
\nonumber\\
& + & |A(B^0 \to [D^{*-}\rho^+]_\parallel)|^2\,|A([D^{*-}\rho^+]_\parallel \to D_s^- K^+)|^2~.
\eea 
Comparing this expression for $|A(B^0 \to D^-_s K^+)|^2$ with that given in (\ref{effective1})
and (\ref{+}), we find
\beq
|A([D^{*-}\rho^+]_+ \to D_s^- K^+)|^2 = g_0 |A([D^{*-}\rho^+]_0 \to D_s^- K^+)|^2
+ g_\parallel |A([D^{*-}\rho^+]_\parallel \to D_s^- K^+)|^2
\eeq
where $g_0=0.924$ and $g_\parallel = 0.076$ are longitudinal and parallel fractions of 
$B^0 \to D^{*-}\rho^+$ decays relative to decays with positive parity. [See comment (a) 
above.]. That is, the effective rescattering probability for positive parity is given by a 
weighted average of the two rescattering probabilities for longitudinal and parallel 
helicity states. 

To conclude this section let us show that using merely time-reversal
invariance and assuming a dominant intermediate state for rescattering  
permits predicting a ratio of $E$ and $T$-type amplitudes for one pair of 
processes in terms of a similar (sometimes given) ratio of another pair of processes. 
Applying relations similar to (\ref{effective1}) and (\ref{effective2}) to VV amplitudes
for a given helicity $h$, one has
 \bea
 A(B^0 \to [D^{*-}_s K^{*+}]_h) & = & A(B^0 \to [D^{*-}\rho^+]_h)\,A([D^{*-}\rho^+]_h 
 \to [D_s^{*-} K^{*+}]_h)~,
 \nonumber \\
 A(B_s \to [D^{*-} \rho^+]_h) & = & A(B_s \to [D_s^{*-} K^{*+}]_h)\,A([D_s^{*-} K^{*+}]_h
 \to [D^{*-} \rho^+]_h)~.
 \eea
Using time-reversal invariance (neglecting the small  $B_s$-$B^0$ mass-difference),
\beq
A([D_s^{*-} K^{*+}]_h \to [D^{*-} \rho^+)]_h) = A([D^{*-}\rho^+]_h \to [D_s^{*-} K^{*+}]_h)~,
\eeq
 one obtains
 \beq
 \frac{A(B_s \to [D^{*-} \rho^+]_h)}{A(B_s \to [D_s^{*-} K^{*+}]_h)} 
 = \frac{A(B^0 \to [D^{*-}_s K^{*+}]_h)}{A(B^0 \to [D^{*-}\rho^+]_h)}~.
 \eeq
Thus a similar relation holds also for ratios of square roots of total branching ratios,
\beq\label{T-ratio}
\sqrt{\frac{\br(B_s \to D^{*-} \rho^+)}{\br(B_s \to D_s^{*-} K^{*+})}} =
\sqrt{\frac{\br(B^0 \to D^{*-}_s K^{*+})}{\br(B^0 \to D^{*-}\rho^+)}}
 = 0.07^{+0.02}_{-0.01}~.
 \eeq
Here we have used $\br(B^0 \to D^{*-}_s K^{*+}) = (3.2^{+1.5}_{-1.3})\times
10^{-5}$ \cite{pdg} and the value of $\br(B^0 \to D^{*-}\rho^+)$ quoted in
Table \ref{tab:BPP}  
for the sum of positive and negative parity states. 
The two ratios of amplitudes in (\ref{T-ratio}), corresponding to values of $|E/T|$ not 
 discussed earlier in our study, lie precisely in the range of $|E/T|$ assumed for all 
 our other predictions.

The relations (\ref{ratio-lambda}) and (\ref{T-ratio}) have been derived for $PP$ 
and $VV$ final states belonging to a class of the pair $(D^- \pi^+, D^-_s K^+)$ 
appearing in the first of Eqs.\,(\ref{BsBPP}). Similar amplitude relations can be 
derived for $PP, VP$ and $VV$ final states belonging to classes of states 
appearing in the other two equations. 
For instance, the rescattering relations,
\bea
A(B^0 \to D^{*+}_s D^-_s) & = & A(B^0 \to D^{*+} D^-)
A(D^{*+} D^-\to D^{*+}_s D^-_s)~, \nonumber \\
A(B_s \to D^{*+} D^-) & = & A(B_s \to D^{*+}_s D^-_s)
A(D^{*+}_s D^-_s \to D^{*+} D^-)~,
\eea
and time-reversal invariance,
\beq
A(D^{*+} D^-\to D^{*+}_s D^-_s) = A(D^{*+}_s D^-_s \to D^{*+} D^-)~,
\eeq
imply
\beq\label{ratio1}
\frac{A(B_s \to D^{*+} D^-)}{A(B_s \to D^{*+}_s D^-_s)}
= \frac{A(B^0 \to D^{*+}_s D^-_s)}{A(B^0 \to D^{*+} D^-)}~.
\eeq
Similarly, the relations
\bea
A(B^0 \to D_s^+K^{*-}) &= & A(B^0 \to D^+ \rho^-)\,A(D^+ \rho^-\to D_s^+ K^{*-})~,
\nonumber \\
A(B_s \to D^+ \rho^-) & = & A(B_s \to D_s^+ K^{*-})\,A(D_s^+ K^{*-} \to D^+ \rho^-)~,
\eea
and invariance of rescattering under time-reversal lead to
\beq\label{ratio2}
\frac{A(B_s \to D^+ \rho^-)}{ A(B_s \to D_s^+ K^{*-})} =
\frac{A(B^0 \to D_s^+K^{*-})}{A(B^0 \to D^+ \rho^-)}~.
\eeq
While experimental information exists on $T$-type amplitudes in the two denominators
in (\ref{ratio1}) (see Table \ref{tab:predVP}), the four numerators in this equation
and in (\ref{ratio2}) representing $E$-type amplitudes have not yet been measured. 
We expect the magnitudes of all four $|E/T|$ ratios to lie in the range $0.07 \pm 0.02$.

\section{Summary and conclusions
\label{sec:sum}}

We have shown that some observed $B$ decays which have been cited as evidence 
for exchange and annihilation processes can be generated by rescattering from
decays whose amplitudes do not involve the spectator quark and hence are
not suppressed by powers of $f_B/m_B$.  We have studied a number of processes
such as $B^0 \to K^+ K^-$, $B_s \to \pi^+ \pi^-$, and $B^+ \to D_s^+ \phi$,
and have identified promising states from which they can be generated by
rescattering.  We have found that such decays have typical
amplitude ratios ranging from 5\% to 20\% with respect to the largest amplitude from
which they can rescatter.  

Using a narrower range between 5\% and 10\% associated with exchange amplitudes, 
we have predicted branching fractions, in a vast range from
${\cal O}(10^{-9})$ to ${\cal O}(10^{-4})$
for a large number of as-yet-unseen $B$ and $B_s$ 
decay processes. These include $B^0$ decays to $K^+ K^-, K^{*\pm}K^{*\mp}$, 
$D^{(*)+}_s D^{(*)-}_s$, 
$D^{(*)0} \bar D^{(*)0}, D^{(*)+} D^{(*)-}$ and $B_s$ decays to 
$D^{(*)\pm} D^{(*)\mp}$, $D^{(*)0}\bar D^{(*)0}$, $D^\pm \pi^\mp$, 
$D^0(\bar D^0)\pi^0$. Predictions of order a few times $10^{-7}$ have also 
been presented for $\br(B^+ \to D^+ K^0)$ and $\br(B_s \to \rho^+\pi^-), 
\br(B_s \to \rho^-\pi^+)$, providing tests for the suppression of annihilation and 
penguin annihilation amplitudes. Other predictions for $\br(B_s \to D^{*+}\pi^-)$ 
and $\br(B_s \to D^-\rho^+)$ 
around $(1-2)\cdot 10^{-6}$ have been derived in the 
limit of U-spin symmetry. Finally, a class of processes has been identified in 
which time-reversal 
invariance of strong interactions leads to further relations between ratios 
of exchange amplitudes
and unsuppressed amplitudes.

{\bf Note added}: After the completion of this work we were made aware of an 
unpublished measurement of $\br(B_s \to D^+ D^-)$ by the LHCb collaboration 
\cite{LHCb}, $\br(B_s \to D^+ D^-)/\br(B^0 \to D^+ D^-) = 1.00 \pm 0.18 \pm
0.09$. Using the values $\br(B^0 \to D^+ D^-)$ and $f_+\br(B_s \to D_s^{*+}
D_s^{*-})$ in Tables \ref{tab:BPP} and \ref{tab:BsPP} we calculate for $B_s
\to D^+ D^-$ a ratio $|E/T| = 0.11 \pm 0.02$, on the high side of the range
which we have assumed.

\section*{Acknowledgements}

We would like to thank Dan Pirjol for useful discussions and Tim Gershon for
informing us about 
Refs.\,\cite{CKMfitter} and \cite{LHCb}.
This work was supported in part by NSERC of Canada (DL) and by the United States
Department of Energy through Grant No.\ DE FG02 90ER40560 (JLR).

\end{document}